\documentclass[conference]{IEEEtran}
\IEEEoverridecommandlockouts

\usepackage{amsmath}
\usepackage{amssymb}
\usepackage{amsfonts}
\usepackage{algorithmic}
\usepackage{hyperref}
\usepackage{textcomp}
\usepackage{xcolor}

\usepackage{listings}
\usepackage{array}
\usepackage{booktabs}

\UseRawInputEncoding

\usepackage[labelfont=bf]{caption}

\usepackage{epstopdf}
\usepackage{comment}
\usepackage{textcase}

\usepackage{pgfplots}
\usepackage{pgfplotstable}
\pgfplotsset{compat=newest}
\pgfplotsset{compat=1.18}
\usetikzlibrary{3d}
\usepackage{eqparbox}
\usepackage{hyperref}
\usepackage{caption}
\usepackage{tikz}
\usepackage{tikz-3dplot}
\usetikzlibrary{positioning}
\usetikzlibrary{shapes.geometric, calc}
\usetikzlibrary{shapes.geometric, arrows}
\usetikzlibrary{shapes.geometric, arrows.meta}

\usetikzlibrary{positioning, shapes.geometric, arrows}
\tikzstyle{block} = [rectangle, rounded corners, minimum width=2.2cm, minimum height=1cm,text centered, draw=black, fill=cyan!30]
\tikzstyle{arrow} = [thick,->,>=stealth, color=red]

\tikzstyle{block} = [rectangle, rounded corners, minimum width=1.4cm, minimum height=0.6cm, text centered, draw=black, fill=cyan!30]
\tikzstyle{arrow} = [thick, ->, >=stealth, color=blue]

\usepackage{listings}

\tikzstyle{process} = [rectangle, rounded corners, minimum width=3cm, minimum height=1cm, text centered, draw=black, fill=blue!20]
\tikzstyle{arrow} = [thick,->,>=stealth]
\tikzstyle{process} = [rectangle, rounded corners, minimum width=2.5cm, minimum height=1cm, text centered, draw=black, fill=blue!20, text width=2.5cm]
\tikzstyle{arrow} = [thick,->,>=stealth]
\tikzstyle{process} = [rectangle, rounded corners, minimum width=2.5cm, minimum height=1cm, text centered, draw=black, fill=blue!20, text width=2.5cm]
\tikzstyle{arrow} = [thick,->,>=stealth]

\begin{document}
\title{Evolution of IVR building techniques: from code writing to AI-powered automation}

\author{
\IEEEauthorblockN{Khushbu Mehboob Shaikh}
\IEEEauthorblockA{\textit{Technical Lead, Staff Technical Account Manager} \\
\textit{Twilio Inc.}\\
Irving, Texas, United States \\
ORCID: 0009-0000-8681-5830
}
\and
\IEEEauthorblockN{Georgios Giannakopoulos}
\IEEEauthorblockA{\textit{Independent Researcher} \\
The Hague, The Netherlands \\
ORCID: 0000-0002-3707-3276}
}
\maketitle
\begin{abstract}
Interactive Voice Response (IVR) systems have undergone significant transformation in recent years, moving from traditional code-based development to more user-friendly approaches leveraging widgets and, most recently, harnessing the power of Artificial Intelligence (AI) for automated IVR flow creation. This paper explores the evolution of IVR building techniques, highlighting the industry's revolution and shaping the future of IVR systems. The authors delve into the historical context, current trends, and future prospects of IVR development, elucidating the impact of AI on simplifying IVR creation processes and enhancing customer experiences.
\end{abstract}
\begin{IEEEkeywords}
Interactive Voice Response, AI-Powered Automation, Machine Learning, Customer Experience, Widget-Based Development, Call Center Operations, Natural Language Processing
\end{IEEEkeywords}

\section{Introduction}
Interactive Voice Response (IVR) systems have become integral to modern call center operations, serving as the first point of contact between customers and businesses. These systems enable automated interactions, allowing customers to navigate menus, access information, and perform tasks such as paying bills or checking account balances without the need for a human operator \cite{b1},\cite{b2}. The efficiency and cost-effectiveness of IVR systems make them a vital tool for managing high volumes of customer inquiries and transactions.

Traditionally, IVR systems were developed using complex, code-based approaches. This involved writing extensive scripts and logic to handle various call scenarios, which required significant programming expertise and resources \cite{b3}. Developers had to manually code the entire IVR flow, including menu options, prompts, and routing rules \cite{b4}. Maintaining and updating these systems was labor-intensive, often necessitating a deep understanding of both the underlying code and the specific requirements of the business. As a result, traditional IVR development was not only time-consuming but also prone to errors and inefficiencies.

Recognizing these challenges, the industry began to shift towards more user-friendly development methods. The emergence of widget-based platforms marked a significant step forward, allowing non-technical users to design IVR systems through graphical interfaces \cite{b2},\cite{b5}. These platforms provided drag-and-drop functionality, enabling the creation and modification of IVR flows without the need for detailed coding knowledge. This democratization of IVR development significantly reduced the time and cost associated with implementing and maintaining these systems.

The most recent and transformative advancement in IVR technology is the integration of Artificial Intelligence (AI)\cite{b6}. AI-powered automation has revolutionized the way IVR systems are developed and operated\cite{b7}. By leveraging natural language processing (NLP) and machine learning algorithms, AI can automate the creation of IVR flows, making them more intuitive and responsive to customer needs. This advancement not only enhances the efficiency of IVR systems but also improves the overall customer experience by providing more personalized and accurate interactions\cite{b8}.

\section{EVOLUTION OF IVR BUILDING TECHNIQUES}
\subsection{Traditional Code-Based IVR Development}
In the early days of IVR technology, creating interactive voice response systems was a highly complex and technical process\cite{b1}. Developers needed to write extensive amounts of code to define every aspect of the call flow, from greeting messages and menu options to call routing and data retrieval\cite{b9}. This traditional code-based development required deep expertise in programming languages and telecommunications protocols.

The process began with understanding the business requirements and translating them into detailed scripts that would guide the interaction between the IVR system and the caller. Each possible caller action had to be anticipated and coded accordingly, resulting in large, intricate codebases that defined the entire call flow.

Writing these scripts was labor-intensive and time-consuming. Developers needed to ensure that the logic for handling various scenarios was accurate and that the system could handle different types of user inputs correctly. Any change or update to the IVR system required significant effort, as it involved modifying the code, testing the changes, and deploying the updated system. This often meant that even minor adjustments could take considerable time and resources.

Due to the complexity of the code and the need for precise logic, developing IVR systems was typically reserved for individuals with substantial technical expertise. This reliance on highly skilled developers limited the accessibility of IVR technology to organizations that could afford such specialized talent. Moreover, the technical nature of the development process made it difficult for business stakeholders to directly influence or modify the IVR system, often leading to communication gaps and delays in implementation.

Maintaining and updating traditional code-based IVR systems posed significant challenges. As business requirements evolved, the IVR system needed to be updated to reflect new processes, services, or regulations. Each update carried the risk of introducing errors, as even a small mistake in the code could disrupt the entire call flow. This risk was particularly pronounced in larger IVR systems with extensive codebases, where identifying and fixing issues could be a daunting task. The need for rigorous testing and validation further added to the complexity and time required for maintenance.

Scaling traditional IVR systems to handle increased call volumes or additional features also proved challenging \cite{b1}. Adding new functionalities often involved writing more code and integrating it seamlessly with the existing system. This could lead to longer development cycles and higher costs, making it difficult for businesses to quickly adapt to changing customer needs or market conditions.

In summary, the traditional code-based approach to IVR development was characterized by its labor-intensive nature, the necessity for deep technical knowledge, and significant challenges in maintenance and scalability. While it laid the foundation for automated customer interactions, the limitations and complexities of this method underscored the need for more efficient and user-friendly IVR development techniques.

\subsection{Emergence of Widget-Based Development}
As the limitations of traditional code-based IVR development became apparent, the industry began to seek more user-friendly solutions. This quest for simplification led to the emergence of widget-based development platforms \cite{b2}, which revolutionized the way IVR systems were designed and implemented.

Widget-based platforms introduced graphical user interfaces (GUIs) that allowed users to design IVR call flows visually. Instead of writing lines of code, users could drag and drop pre-built widgets or modules to create their IVR systems. These widgets represented different components of the IVR, such as menu options, call transfers, and data inputs. This visual approach made the development process intuitive and accessible, even for those without technical expertise.

One of the most significant advantages of widget-based platforms was their accessibility. Business analysts, customer service managers, and other non-technical users could now participate directly in the IVR design process. They could create, modify, and optimize IVR systems without needing to rely on developers. This democratization of IVR development empowered a broader range of stakeholders to contribute to the creation and refinement of customer interaction flows.

By simplifying the design process, widget-based platforms dramatically reduced the time required to develop and deploy IVR systems. Changes that previously took days or weeks to implement could now be completed in a matter of hours. This agility allowed businesses to respond more quickly to evolving customer needs, market conditions, and regulatory requirements. Rapid prototyping and iteration became feasible, enabling continuous improvement of IVR systems.

The visual nature of widget-based platforms also facilitated better collaboration between technical and non-technical teams. Business requirements could be translated directly into IVR designs without the risk of miscommunication or misinterpretation. This alignment between business objectives and technical implementation led to more effective and efficient IVR systems that better met customer expectations.

Maintaining and updating IVR systems became more straightforward with widget-based platforms. Users could easily modify call flows, add new features, or make adjustments through the GUI. This reduced the risk of errors and minimized the need for extensive testing and validation. Additionally, the modular nature of widgets allowed for scalable designs, enabling businesses to expand their IVR systems as needed without significant redevelopment.

In conclusion, the emergence of widget-based development platforms marked a significant milestone in the evolution of IVR building techniques. By making the development process more accessible, reducing the time required for implementation, and improving collaboration between teams, these platforms addressed many of the challenges associated with traditional code-based methods \cite{b2}. This shift not only enhanced the efficiency and effectiveness of IVR systems but also paved the way for further innovations in IVR development.

\section{THE RISE OF AI IN IVR DEVELOPMENT}
With advancements in AI, IVR systems have entered a new era. AI technologies, such as natural language processing (NLP) and machine learning algorithms, automate the creation of IVR flows. This section explores how AI has revolutionized IVR development, offering benefits like improved efficiency, personalized interactions, and enhanced customer experiences.

\subsection{Automation of IVR Creation}
\begin{figure}[htbp]
\centerline{\includegraphics[width=3.5in]{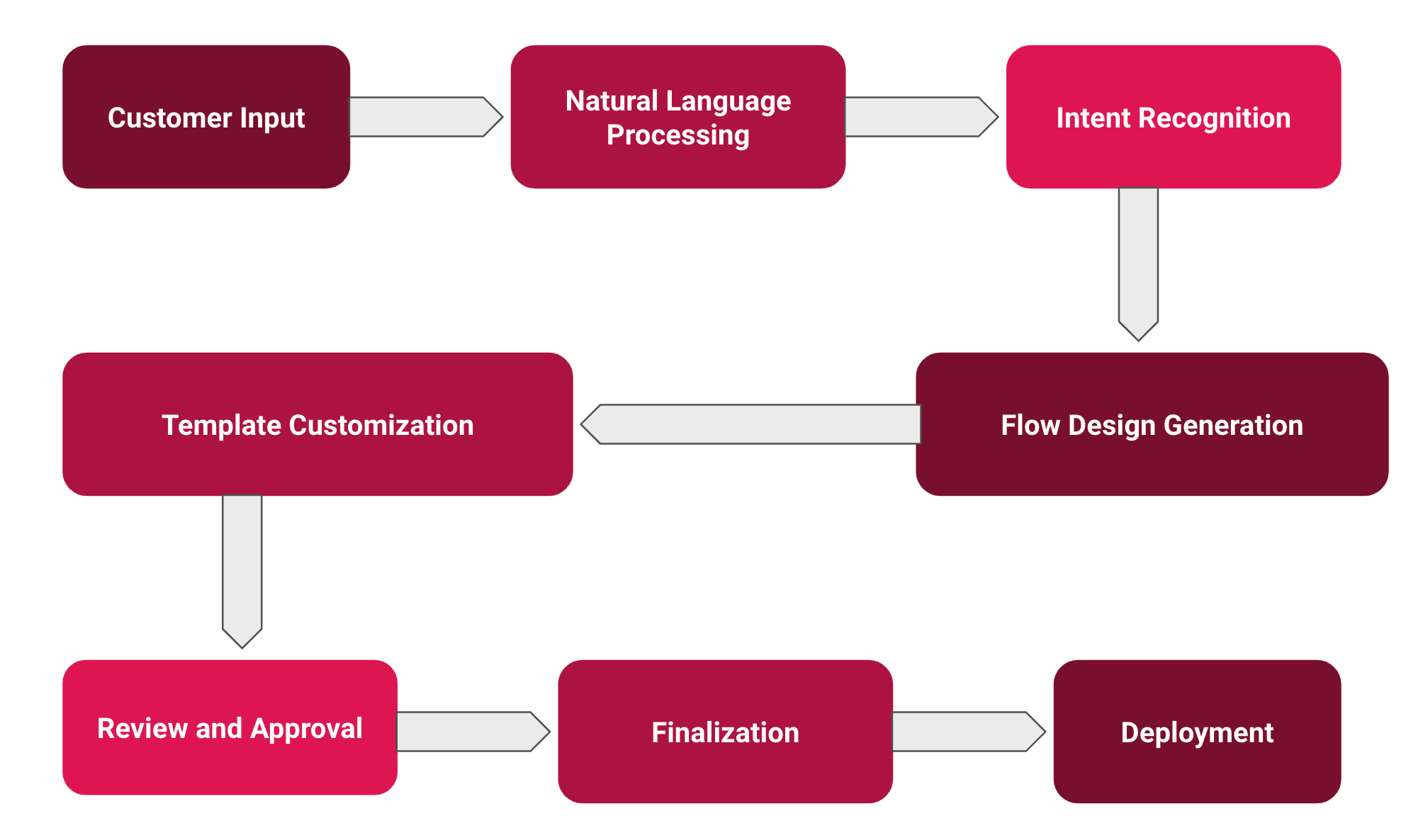}}
\caption{A flow diagram that illustrates how AI can take a prompt from a customer and build an IVR flow template.}
\label{fig1}
\end{figure}

AI has significantly transformed the process of developing IVR systems by introducing automation. Traditional methods required extensive manual effort to design and implement call flows, which was both time-consuming and prone to errors. AI-driven tools, however, can generate IVR flows automatically by leveraging vast amounts of data and sophisticated algorithms. These tools analyze customer interactions, identify common patterns, and create optimized call flows that address user needs more effectively. \hyperref[fig1]{Figure 1} demonstrates how AI can take a customer prompt and build an IVR flow template, which can then be customized according to customer requirements before deployment.

\subsection{Natural Language Processing (NLP)} 
One of the most groundbreaking advancements in AI-powered IVR systems is the integration of NLP. Unlike conventional IVR systems that relied on predefined menu options and rigid structures, NLP enables IVR systems to understand and interpret human language in a more natural and intuitive manner. This allows customers to interact with the system using their own words and phrases rather than following a strict set of commands. NLP technology can decipher intent, context, and sentiment, enabling more fluid and efficient conversations between customers and the IVR system \cite{b5}.

\subsection{Machine Learning Algorithms}  
Machine learning algorithms play a crucial role in enhancing the capabilities of AI-powered IVR systems. These algorithms continuously learn from interactions, improving their accuracy and performance over time \cite{b6}. By analyzing historical call data, machine learning models can predict customer needs and preferences, allowing the IVR system to provide more relevant and timely responses. This adaptability ensures that the system remains effective even as customer behavior and expectations evolve.

\subsection{Personalized Customer Interactions}  
AI-powered IVR systems excel at delivering personalized customer interactions \cite{b5}. By leveraging customer data and insights, these systems can tailor responses to individual preferences and histories. For example, a returning customer might be greeted by name and offered assistance based on their previous interactions with the company. This level of personalization enhances the overall customer experience, making interactions more engaging and satisfying.

\subsection{Enhanced Efficiency} 
AI-driven automation significantly boosts the efficiency of IVR systems \cite{b4}. Traditional systems required manual updates and maintenance, which could be time-consuming and resource-intensive. In contrast, AI-powered IVR systems can autonomously adapt to changing conditions, reducing the need for constant human intervention. This leads to faster response times, shorter call durations, and increased overall efficiency. Additionally, the ability to handle complex queries and provide accurate information reduces the need for escalation to human agents, further streamlining operations \cite{b9}.

\subsection{Improved Customer Experiences}  
The ultimate goal of AI-powered IVR systems is to enhance customer experiences. By offering more natural, intuitive, and personalized interactions, these systems address common pain points associated with traditional IVR systems, such as frustration with rigid menus and lengthy wait times. Customers benefit from quicker resolutions, more relevant information, and a smoother overall experience \cite{b7}. The ability to handle a wide range of inquiries and provide accurate responses also contributes to higher levels of customer satisfaction and loyalty.

In summary, the rise of AI in IVR development has revolutionized the way these systems are designed and implemented. Through automation, natural language processing, and machine learning, AI-powered IVR systems offer improved efficiency, personalized interactions, and enhanced customer experiences. These advancements have not only addressed the limitations of traditional IVR systems but also set new standards for customer service and engagement in the digital age.

\section{INDUSTRY IMPACT AND FUTURE OUTLOOK}
The integration of AI into IVR systems has brought about profound changes in the industry, leading to increased adoption rates, substantial cost savings, and higher levels of customer satisfaction \cite{b4}. This section delves into the broader impact of AI-powered IVR systems on the industry and explores emerging trends that promise to further revolutionize these systems, making them even more intuitive and effective.

\subsection{Increased Adoption Rates} 
\begin{figure}[htbp]
\centerline{\includegraphics[width=3.5in]{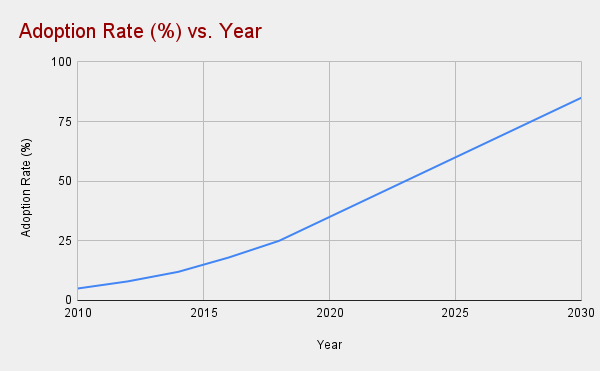}}
\caption{Adoption rates of AI in IVR systems.}
\label{fig2}
\end{figure}

The advancements in AI technology have made IVR systems more accessible and appealing to a wider range of businesses. Traditional IVR systems required significant investment in technical expertise and resources, limiting their adoption primarily to large enterprises. However, AI-powered IVR solutions are more user-friendly and cost-effective, allowing small and medium-sized businesses to implement and benefit from them. This democratization of IVR technology has led to a surge in adoption rates across various industries. \hyperref[fig2]{Figure 2} illustrates the significant growth in the adoption rate of AI in IVR systems year over year. The data used to depict this graph is hypothetical and intended for illustrative purposes only.

\subsection{Cost Savings} 
One of the most significant advantages of AI-integrated IVR systems is the potential for substantial cost savings. By automating routine inquiries and tasks, these systems reduce the need for large customer service teams, thereby lowering operational costs. AI-powered IVR systems can handle high volumes of calls efficiently, minimizing the burden on human agents and reducing the costs associated with training, hiring, and managing a large workforce. Additionally, the ability to provide accurate and timely responses decreases call duration and improves first-call resolution rates, further contributing to cost savings.

\subsection{Higher Customer Satisfaction}  
\begin{figure}[htbp]
\centerline{\includegraphics[width=3.5in]{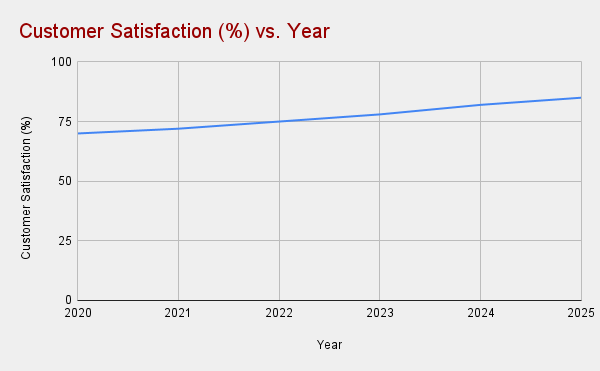}}
\caption{Growth in customer satisfaction.}
\label{fig3}
\end{figure}

AI-powered IVR systems have significantly improved the customer experience, leading to higher satisfaction levels \cite{b9}. The use of natural language processing (NLP) allows customers to interact with the system more naturally, avoiding the frustration often associated with rigid menu options and long wait times. Personalized interactions and quicker resolutions enhance the overall experience, making customers feel valued and understood. As a result, businesses that adopt AI-powered IVR systems often see an increase in customer loyalty and retention. 
\hyperref[fig3]{Figure 3} shows the trend in customer satisfaction over the years. The data represented in this graph is hypothetical and intended for illustrative purposes only.

\subsection{Voice Biometrics}  
Voice biometrics is an emerging trend that promises to add a new layer of security and personalization to IVR systems \cite{b5}. By analyzing unique vocal characteristics, voice biometrics can accurately verify a caller's identity, reducing the need for cumbersome security questions and PINs. This technology not only enhances security but also streamlines the authentication process, providing a smoother and more efficient experience for customers. As voice biometrics becomes more advanced and widely adopted, it is expected to become a standard feature in AI-powered IVR systems.

\subsection{Sentiment Analysis}   
Sentiment analysis is another promising development in the evolution of IVR systems. By leveraging natural language processing (NLP) and machine learning algorithms, sentiment analysis can detect the emotional tone of a caller's voice \cite{b3}. This capability allows IVR systems to identify frustrated or dissatisfied customers and prioritize their calls for escalation to human agents. By addressing negative sentiments proactively, businesses can improve customer satisfaction and prevent potential issues from escalating. Sentiment analysis also provides valuable insights into customer emotions, helping businesses refine their communication strategies and service offerings.

\subsection{Predictive Analytics}  
Predictive analytics is transforming the way IVR systems anticipate and respond to customer needs. By analyzing historical data and identifying patterns, predictive analytics can forecast customer inquiries and issues before they occur. This enables IVR systems to provide proactive support, such as pre-emptively addressing common problems or offering tailored recommendations. Predictive analytics also helps businesses optimize their resources by forecasting call volumes and adjusting staffing levels accordingly. Based on industry reports and academic studies on predictive analytics, the ongoing advancements in IVR technology are expected to significantly enhance the effectiveness and efficiency of AI-powered IVR systems.

\subsection{Future Prospects}   
The future of IVR systems is bright, with ongoing advancements in AI and related technologies paving the way for even more sophisticated and intuitive solutions. Emerging trends such as conversational AI \cite{b10}, which enables more natural and human-like interactions, and the integration of AI with other communication channels, such as chatbots and messaging platforms, will further enhance the capabilities of IVR systems. Additionally, the continuous improvement of machine learning models will enable IVR systems to handle increasingly complex inquiries and provide more accurate responses.

In summary, the integration of AI into IVR systems has had a transformative impact on the industry, driving increased adoption rates, cost savings, and higher customer satisfaction. Emerging trends like voice biometrics, sentiment analysis, and predictive analytics promise to further revolutionize IVR systems, making them more secure, intuitive, and effective. As businesses continue to embrace AI-powered IVR solutions, the future holds exciting possibilities for enhancing customer service and communication.

\section{CHALLENGES AND CONSIDERATIONS}
Despite the numerous benefits and advancements brought by AI-powered IVR development, there are several challenges and considerations that organizations must address to fully leverage these technologies effectively.

\subsection{Data Privacy Concerns}  
One of the foremost challenges in deploying AI-powered IVR systems is ensuring data privacy and security. IVR systems often handle sensitive customer information, including personal identification details and financial data. The integration of AI introduces additional data touchpoints, increasing the potential for data breaches or misuse. Organizations must implement robust data encryption, comply with regulatory standards such as GDPR or CCPA, and establish stringent access controls to protect customer data \cite{b5}.

\subsection{Integration Complexities} 
Integrating AI-powered IVR systems with existing infrastructure can be complex. Legacy systems may not be designed to support modern AI technologies, necessitating significant overhauls or the implementation of middleware to facilitate compatibility. Seamless integration requires careful planning, a deep understanding of both the existing and new systems, and often, substantial investment in time and resources to ensure that the AI components work harmoniously with existing business processes.

\subsection{Continuous Training and Optimization} 
AI models rely heavily on the quality and quantity of data they are trained on. To maintain their accuracy and relevance, these models require continuous training and updating \cite{b6}. This ongoing optimization ensures that the IVR system can adapt to new patterns in customer interactions and emerging trends. Organizations must allocate resources for regular data collection, model retraining, and performance monitoring to keep their AI-powered IVR systems performing at optimal levels.

\subsection{Balancing Automation with Human Touch}  
While AI can handle a significant portion of customer interactions, there will always be situations where human intervention is necessary. Striking the right balance between automation and the human touch is crucial. Over-reliance on AI can lead to impersonal customer experiences, while insufficient automation can negate the efficiency gains. Organizations need to design their IVR systems in a way that ensures complex or sensitive issues are escalated to human agents promptly, preserving the quality of customer service.

\subsection{Managing Customer Expectations} 
AI-powered IVR systems can significantly enhance customer experiences, but they also raise expectations. Customers may anticipate faster, more accurate responses, and personalized interactions. Any shortcomings or errors in the AI system can lead to customer dissatisfaction \cite{b9}. Managing these expectations through clear communication and setting realistic service standards is essential to maintaining customer trust and satisfaction.

In summary, while AI-powered IVR systems offer transformative potential, addressing these challenges is critical to unlocking their full benefits. By focusing on data privacy, integration, continuous optimization, balancing automation with human elements, and managing customer expectations, organizations can harness the power of AI to create more efficient, effective, and satisfying IVR experiences.

\section{CONCLUSION}
The evolution of IVR building techniques from manual code writing to AI-powered automation marks a significant advancement in the industry. This transformation reflects the broader trend towards leveraging cutting-edge technology to enhance operational efficiency and customer experience.

\subsection{Historical Perspective} 
Initially, IVR systems relied heavily on manual code writing, which was labor-intensive and required deep technical expertise. This traditional approach, while functional, was fraught with challenges such as high maintenance costs, difficulty in updating call flows, and susceptibility to errors. These limitations often resulted in a rigid, less responsive customer service environment.

\subsection{Transition to Widget-Based Development}  
The emergence of widget-based development platforms represented a pivotal shift. These platforms made IVR system creation accessible by offering graphical interfaces that enabled non-technical users to easily design and modify call flows. This shift not only reduced the complexity and time required to deploy IVR solutions but also empowered a broader range of users to contribute to the development process. The accessibility and ease-of-use of widget-based systems marked a significant improvement in the agility and responsiveness of IVR systems.

\subsection{The Advent of AI-Powered Automation}  
The most recent and transformative advancement in IVR development is the integration of AI technologies. AI-powered automation, driven by natural language processing (NLP) and machine learning algorithms, has revolutionized the way IVR systems are built and operated. These technologies enable the creation of more sophisticated and dynamic IVR flows that can adapt to user inputs in real-time, providing more personalized and efficient customer interactions \cite{b10}.

\subsection{Benefits of AI Integration} 
AI-powered IVR systems offer numerous benefits, including improved operational efficiency, reduced costs, and enhanced customer satisfaction. By automating routine tasks and allowing for more complex and natural interactions, AI enhances the overall user experience. Customers can navigate IVR systems more intuitively, and organizations can handle higher call volumes with greater accuracy and speed. Additionally, AI enables continuous learning and improvement, allowing IVR systems to evolve and improve over time.

\subsection{The Importance of Embracing AI}  
In today's rapidly evolving technological landscape, embracing AI-driven advancements in IVR systems is crucial for organizations to remain competitive. The ability to provide superior customer service through efficient and intelligent IVR systems can be a key differentiator in the marketplace. Organizations that leverage AI to enhance their IVR systems will be better positioned to meet the growing expectations of their customers and respond to the dynamic demands of the market.

\subsection{Looking Forward}   
The future of IVR development promises even greater integration of advanced technologies, such as voice biometrics, sentiment analysis, and predictive analytics. These emerging trends will further enhance the capabilities of IVR systems, making them even more intuitive, secure, and effective. Organizations must stay abreast of these developments and continuously innovate to maintain a competitive edge in customer service and communication.

In conclusion, the journey from manual code-based IVR development to AI-powered automation underscores the significant strides made in enhancing IVR systems. By embracing these advancements, organizations can ensure they are well-equipped to provide exceptional customer service and thrive in the ever-evolving landscape of customer communication. The future of IVR systems lies in continuous innovation and the strategic application of AI technologies to meet and exceed customer expectations.

\section*{Acknowledgment}
The authors would like to express their gratitude to the AI-powered tools and resources that significantly contributed to the completion of this research paper. The data for the graphs and insights on the impact and advancements of AI in Interactive Voice Response (IVR) technology were greatly facilitated by the use of these AI tools.

Additionally, the authors would like to acknowledge the support from various industry reports and academic studies on predictive analytics and AI integration in communication technologies. These sources have enriched the content and provided a robust foundation for the analysis presented in this paper.

The authors are grateful to researchers, developers, and organizations that continue to push the boundaries of AI and IVR technologies, thereby enabling the creation of more efficient and intuitive systems.

\bibliographystyle{IEEEtran}
\bibliography{anystyle}

\end{document}